\documentclass{emulateapj}
\usepackage{mathptmx} 

\usepackage{amssymb,amsmath}
\usepackage[hidelinks]{hyperref}
\usepackage[varg]{txfonts} 

\shorttitle{Kiloparsec-scale jets in radio-loud NLS1s}

\shortauthors{J.~L. Richards \& M.~L. Lister}

\newcommand{\sourcename}[1]{\mbox{#1}} 

\begin{document}

\title{Kiloparsec-scale jets in three radio-loud narrow-line Seyfert~1 galaxies}

\author{Joseph~L. Richards$^*$}
\author{Matthew~L. Lister}
\email[$^*$]{jlr@purdue.edu}
\affil{Department of Physics and Astronomy, Purdue University, 525 Northwestern Ave, West Lafayette, IN 47907, USA}

\begin{abstract}\vspace*{3ex}
  We have discovered kiloparsec-scale extended radio emission in three
  narrow-line Seyfert~1 galaxies (NLS1s) in sub-arcsecond resolution
  9~GHz images from the Karl G. Jansky Very Large Array (VLA). We find
  all sources show two-sided, mildly core-dominated jet structures
  with diffuse lobes dominated by termination hotspots. These span
  20--70~kpc with morphologies reminiscent of FR~II radio galaxies,
  while the extended radio luminosities are intermediate between FR~I
  and FR~II sources. In two cases the structure is linear, while a
  45\degr{} bend is apparent in the third. Very Long Baseline Array
  images at 7.6~GHz reveal parsec-scale jet structures, in two cases
  with extended structure aligned with the inner regions of the
  kiloparsec-scale jets. Based on this alignment, the ratio of the
  radio core luminosity to the optical luminosity, the jet/counter-jet
  intensity and extension length ratios, and moderate core brightness
  temperatures ($\lesssim10^{10}$~K), we conclude these jets are
  mildly relativistic ($\beta\lesssim0.3$, $\delta\sim1-1.5$) and
  aligned at moderately small angles to the line of sight
  (10--15\degr{}). The derived kinematic ages of
  $\sim10^{6}$--$10^{7}$~y are much younger than radio galaxies but
  comparable to other NLS1s. Our results increase the number of
  radio-loud NLS1s with known kiloparsec-scale extensions from seven
  to ten and suggest that such extended emission may be common, at
  least among the brightest of these sources.
\end{abstract}



\section{Introduction}
%
Narrow-line Seyfert~1 galaxies (NLS1s) are a peculiar class of active
galactic nuclei (AGN). They resemble ordinary Seyfert~1 galaxies but
their optical broad lines are narrower (${\mathrm{FWHM}(H\betaup)\leq
  2000~\mathrm{km\,s^{-1}}}$). The black hole masses of NLS1s are
relatively small~\citep[${10^{5}-10^{8}}$~M$_{\sun}$;
e.g.,][]{zhou_comprehensive_2006} and their accretion rates are high,
often a substantial fraction of the Eddington
rate~\citep{boller_soft_1996}. It has, however, been suggested that
this is a result of systematic underestimation of the black hole
masses~\citep{decarli_are_2008, marconi_effect_2008,
  calderone_black_2013}.  About 7\% of known NLS1s are
radio-loud~\citep{komossa_radio-loud_2006} and it is clear that jets
are present in some of these~\citep[e.g.,][]{zhou_j0948_2003,
  yuan_population_2008}. The detection of gamma-ray emission from
several radio-loud NLS1s provides evidence that these jets are
relativistic~\citep[e.g.,][]{fermi_nls1_class_2009}.  This was
unexpected because jets in AGN are typically associated with large
black holes, while high accretion rates in stellar-mass black hole
binaries normally correspond to quenched-jet
states~\citep[e.g.,][]{Boroson_Black_2002,
  Maccarone_connection_2003}. Study of the jets in NLS1s will
therefore help us understand how the processes underlying jet
production scale with black hole
mass~\citep[e.g.,][]{heinz_sunyaev_2003,
  foschini_nls1_unification_2012}.

Radio emission on kiloparsec scales has been found in only a few
NLS1s. Together, \citet{whalen_optical_2006},
\citet{anton_0324+3410_2008}, \citet{gliozzi_panchromatic_2010}, and
\citet{doi_kiloparsec-scale_2012} reported six examples from the
1.4~GHz Faint Images of the Radio Sky at Twenty centimeters (FIRST)
survey \citep{becker_FIRST_1995} and 4.8~GHz Australian Telescope
Compact Array (ATCA) observations. The resolution of these
observations was limited, however. The 5\arcsec{} FIRST beam only
resolves sources $\gtrsim\!\!2\arcsec{}$, while the ATCA beam was
about a factor of two better.  NLS1s are known out to $z=0.8$ so this
corresponds to $\gtrsim\!\!10$~kpc, leaving a range of kiloparsec
scales largely unexplored. Recently, \citet{doi_fanaroff-riley_2014}
reported extended emission in a radio-quiet NLS1.

In this work, we present the detection of kiloparsec-scale radio
emission in three radio-loud NLS1s (\sourcename{J0953+2836},
\sourcename{J1435+3131}, and \sourcename{J1722+5654}). These are the
first such detections in these sources, all of which were detected but
unresolved by the FIRST survey~\citep{white_catalog_1997}. This
increases the number of radio-loud NLS1s known to be extended on
kiloparsec scales from  seven to ten. We also present
parsec-scale radio images showing bright jet-like structures. Where
necessary, we have assumed a flat $\Lambda$-Cold Dark Matter cosmology
with $\Omega_{\Lambda}=0.7$ and $H_0=70~{\rm km\, s^{-1}\,
  Mpc^{-1}}$. Spectral indices are specified with flux density
$S_\nu\propto\nu^{\alpha}$ at frequency $\nu$. Velocities are quoted
as fractions of the speed of light, $\beta=v/c$.

\section{Observations and Data Reduction}
We observed three radio-loud NLS1s with the Karl G. Jansky Very Large
Array (VLA) and the Very Long Baseline Array (VLBA). The VLA
observations were made to obtain radio coordinates accurate enough for
correlation as part of a VLBA program monitoring the 15 known
radio-loud NLS1s above 0\degr{} declination with archival
centimeter-band flux densities exceeding 30~mJy.~(Richards et al., in
prep.). Among the 15, only these three lacked sufficiently
precise coordinates. Properties of our targets and our refined
coordinates are listed in Table~\ref{table:source_properties}.

The VLA observations were made on 15~Feb~2014 and consisted of a
  single 10~minute snapshot of each source sandwiched between short
observations of a nearby phase calibrator with a known position
(better than 2~mas). A short scan on \sourcename{3C 286} was used to
calibrate the flux density scale, applying the appropriate model to
account for resolved structure. At the time of observation, the VLA
was transitioning from BnA to A configuration, with 27 active antennas
on baselines ranging from 0.3~to 31.5~km. The X-band receiver provided
2~GHz bandwidth centered on 9~GHz, which was sampled with the 8-bit
sampler with 2~s integration time. Requantizer gains were applied and
the data were converted to UVFITS format using
CASA~\citep{CASA_reference}. The data were then calibrated in AIPS
following the standard procedures in the AIPS Cookbook and
deconvolved, self-calibrated, and imaged in
DIFMAP~\citep{shepherd_difmap_1994}. After initial calibration of the
phase calibrators, cycles of phase-only self-calibration and CLEAN
were repeated until a few times the residual root-mean-square (rms)
noise level was reached. Amplitude self-calibration was then applied
to determine the final complex gains. To measure source positions,
the gains derived from the phase calibrators were interpolated
onto the science observations. A circular Gaussian component model of
the core region was fitted to the visibilities and the position of the
brightest component is reported in
Table~\ref{table:source_properties}, accurate to about 0.1 beam, or
about 15~mas. For imaging, the science target data were cleaned and
self-calibrated using the same procedures used for the phase
calibrators.

VLBA observations were made on 8~Feb~2014 using the wideband C-band
receiver configured to provide 256~MHz total bandwidth in 8 sub-bands
equally spaced over the 480~MHz band centered on 7.632~GHz with good
weather at all 10 antennas. The total integration time, listed in
  Table~\ref{table:source_properties}, was split into scans separated
  in time to provide good $(u,v)$ coverage. In each scan, 49~s of
full polarization data were recorded as 2-bit samples for a total
2~Gbps data rate, with science target observations interleaved with
observations of a nearby phase reference. Observations at 4.9, 15.4,
and 23.8~GHz were also carried out, but we report only 7.632~GHz total
intensity results here. Full results from the VLBA program, including
polarization measurements and the other frequencies, will be reported
elsewhere. After correlation with the DiFX
correlator~\citep{deller_difx} using our refined coordinates,
the phase-referenced VLBA data were processed following the AIPS
Cookbook, applying gain solutions from the phase references to the
science targets. A CLEAN model was constructed, then phase-only
self-calibration with a 10 minute solution interval was applied and a
final CLEAN was performed. We verified our flux density scale,
accurate to about 5\%, by comparing several calibrators with
preliminary results from Effelsberg 100~m telescope observations made
by the \emph{Fermi}-GST AGN Multi-frequency Monitoring
Alliance~\citep[F-GAMMA;][]{fuhrmann_fgamma, angelakis_fgamma}.

\begin{deluxetable*}{c c c c c c c c c c}
  \tablecaption{Source properties and refined J2000 coordinates.\label{table:source_properties}}
  \tablehead{
    \colhead{Source} &
    \colhead{NVSS Name} &
    \colhead{$z$} &
    \colhead{$M_{\rm BH}$} &
    \colhead{$S_{1.4}$} &
    \colhead{$M_{\rm V}$} &
    \colhead{R.A.} &
    \colhead{Decl.} &
    \colhead{Phase Reference} &
    \colhead{VLBA Time}\\
    \colhead{} &
    \colhead{} &
    \colhead{} &
    \colhead{($10^{7} M_{\sun}$)} &
    \colhead{(mJy)} &
    \colhead{} &
    \colhead{(h m s)} &
    \colhead{(\degr{} \arcmin{} \arcsec{})} &
    \colhead{VLA / VLBA} &
    \colhead{(min)}\\
    \colhead{(1)} &
    \colhead{(2)} &
    \colhead{(3)} &
    \colhead{(4)} &
    \colhead{(5)} &
    \colhead{(6)} &
    \colhead{(7)} &
    \colhead{(8)} &
    \colhead{(9)} &
    \colhead{(10)}}
  \startdata
  \sourcename{J0953+2836} &
  \sourcename{J095317+283601} &
  0.659 &
  6.3 &
  47.94 &
  $-24.04$ &
  09 53 17.102 &
  28 36 01.559 &
  \sourcename{J0956+2515} / \sourcename{J0954+2639} &
  4.8 \\
  %
  \sourcename{J1435+3131} &
  \sourcename{J143509+313149} &
  0.502 &
  3.2 &
  44.72 &
  $-23.01$ &
  14 35 09.495 &
  31 31 47.864 &
  \sourcename{J1435+3012} / \sourcename{J1422+3223} &
  9.8 \\
  %
  \sourcename{J1722+5654} &
  \sourcename{J172206+565452} &
  0.426 &
  2.5 &
  39.83 &
  $-23.58$ &
  17 22 06.029 &
  56 54 51.696 &
  \sourcename{J1657+5705} / \sourcename{J1727+5510} &
  13.0
  \enddata
  %
  \tablecomments{Columns are as follows: (1)~IAU name (J2000); (2)~corresponding source in the NRAO VLA Sky Survey~\citep[NVSS,][]{condon_nrao_1998}; (3)~redshift from \citet{hewett_improved_2010}; (4)~black hole mass from \citet{yuan_population_2008}; (5)~integrated 1.4~GHz flux density from \citet{white_catalog_1997}; (6)~absolute V-band magnitude computed from Sloan Digital Sky Survey Data Release~10~\citep{sdss_dr10} $g$ and $r$ magnitudes using $V=g-0.52\left(g-r\right)-0.03$, after \citet{jester_sloan_2005} with no correction for galactic extinction; (7)-(8)~new J2000 coordinates determined from our VLA observations with an uncertainty of $\pm15$~mas; (9)~IAU name (J2000) of phase calibrators used for VLA and VLBA observations; (10)~total VLBA integration time.}
\end{deluxetable*}

\section{Results}
\begin{figure}
  \includegraphics[width=\columnwidth]{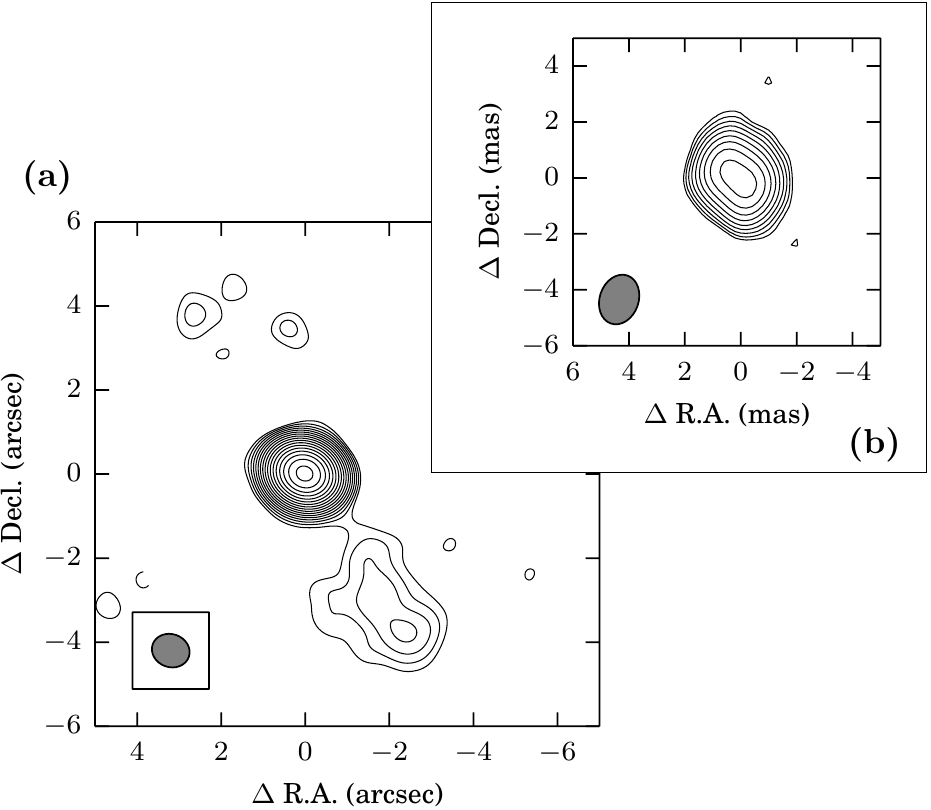}
  \caption{Naturally weighted maps of
    \sourcename{J0953+2836}. Contours increase by factors of
    $\sqrt{2}$. At this redshift, $1\arcsec{}=6.97~\mathrm{kpc}$. (a)
    VLA map (9~GHz). To emphasize low-surface brightness extended
    features, a Gaussian taper with half-maximum at
    ${1.5\times10^{5}}$ wavelengths has been applied. The first
    contour is at $50~\mathrm{\muup Jy\,beam^{-1}}$, about three times
    the map rms of $17~\mathrm{\muup Jy\,beam^{-1}}$, and the map
    peaks at $14.6~\mathrm{mJy\,beam^{-1}}$. The elliptical Gaussian
    restoring beam is shown (FWHM
    $0.911\arcsec\times0.786\arcsec$). (b) VLBA map (7.6~GHz) of the
    central bright feature.  The first contour is at
    $0.3~\mathrm{mJy\,beam^{-1}}$, about three times the map rms of
    $0.1~\mathrm{mJy\,beam^{-1}}$, and the map peaks at
    $5.7~\mathrm{mJy\,beam^{-1}}$. The elliptical Gaussian restoring
    beam is shown (FWHM $1.82~\mathrm{mas}\times1.39~\mathrm{mas}$)}
  \label{fig:j0953+2836}
\end{figure}

\begin{figure}
  \includegraphics[width=\columnwidth]{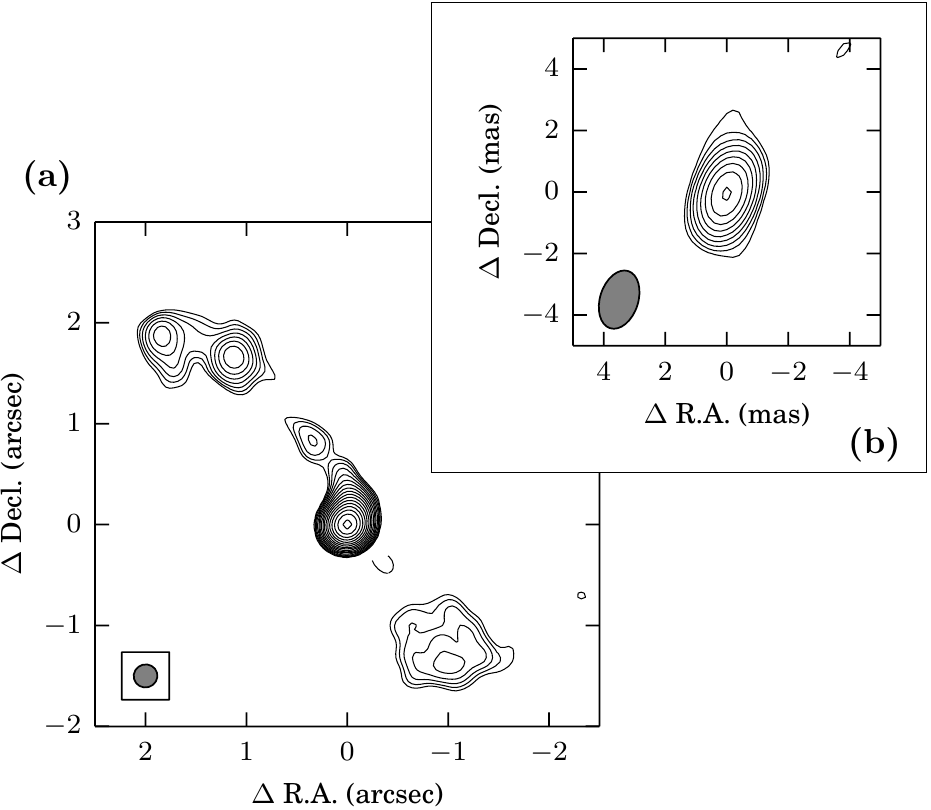}
  \caption{Naturally weighted maps of
    \sourcename{J1435+3131}. Contours increase by factors
    of$\sqrt{2}$. At this redshift,
    $1\arcsec{}=6.12~\mathrm{kpc}$. (a) VLA map (9~GHz). The first
    contour is at $30~\mathrm{\muup Jy\,beam^{-1}}$, about five times
    the map rms noise of $6~\mathrm{\muup Jy\,beam^{-1}}$, and the map
    peaks at $8.39~\mathrm{mJy\,beam^{-1}}$. The elliptical Gaussian
    restoring beam is shown (FWHM $0.236\arcsec\times0.228\arcsec$).
    (b) VLBA map (7.6~GHz) of the central bright feature. The first contour is at
    $0.3~\mathrm{mJy\,beam^{-1}}$, about three times the map rms noise
    of $0.1~\mathrm{mJy\,beam^{-1}}$, and the map peaks at
    $5.0~\mathrm{mJy\,beam^{-1}}$. The elliptical Gaussian restoring
    beam is shown (FWHM $1.95~\mathrm{mas}\times1.24~\mathrm{mas}$).}
  \label{fig:j1435+3131}
\end{figure}

The maps we obtained from our observations are shown in
Figures~\ref{fig:j0953+2836}, \ref{fig:j1435+3131},
and~\ref{fig:j1722+5654} and Table~\ref{table:computed} lists
quantities computed from our results. Using 1.4~GHz FIRST flux
densities and our 9~GHz VLA flux densities, we have determined
the non-simultaneous spectral index $\alpha$. The formal uncertainty
is $\pm0.04$, assuming 5\% uncertainties for the flux densities, not
accounting for non-simultaneity. These spectral indices are lower
limits because our higher resolution maps may resolve out more
extended emission than in the FIRST data.

On kiloparsec scales, all of these sources show similar
morphologies. The emission is dominated by a bright core straddled by
a pair of extended lobes. These lobes are edge-brightened, terminating
in hotspots. Like in FR~II radio galaxies, the ratios of the hotspot
separations to the full extents of the sources exceed
0.5~\citep{fanaroff_riley_1974}. Extended emission between the bright
core and the brighter of the two lobes is found in all sources, while
the fainter lobe is relatively isolated.  On parsec scales, we
  find an isolated core in \sourcename{J1435+3131} while the others
  show extension roughly aligned with the kiloparsec-scale structure.

\subsection{J0953+2836}
The emission in the tapered VLA map of \sourcename{J0953+2836}
(Figure~\ref{fig:j0953+2836}) spans 10\farcs{}1 (measured at the
second contour, following \citet{scheuer_1995}), a projected length of
70.2~kpc. The brighter lobe lies along position angle (PA) 212\degr{},
measured north through east. From the VLBA map, the parsec-scale image
shows extended emission with a ridge along PA 232\degr{}. The ridge
brightens slightly toward the SW. A two-component Gaussian model fits
the visibilities well, with the SW component brighter by a factor of
1.2. Identification of the parsec-scale core is uncertain and it is
unclear whether this is a one- or two-sided jet.

\subsection{J1435+3131}
The kiloparsec-scale emission in \sourcename{J1435+3131}
(Figure~\ref{fig:j1435+3131}) spans 5\farcs{}00, or a projected length
of 30.6~kpc. A sharp bend is evident between the core and the NE
kiloparsec-scale lobe. A narrow, continuous trail of emission exits
northward from the core, then bends about 45\degr{} to the northeast,
terminating in a faint knot. A straight line extended from this knot
along the direction after the bend is closely aligned with the
brighter of two hotspots in the disconnected radio lobe. The
parsec-scale image shows only a compact core. No
kiloparsec-scale emission is evident between the core and the fainter
SW radio lobe, which has a single hotspot.

\subsection{J1722+5654}
\begin{figure}
  \includegraphics[width=\columnwidth]{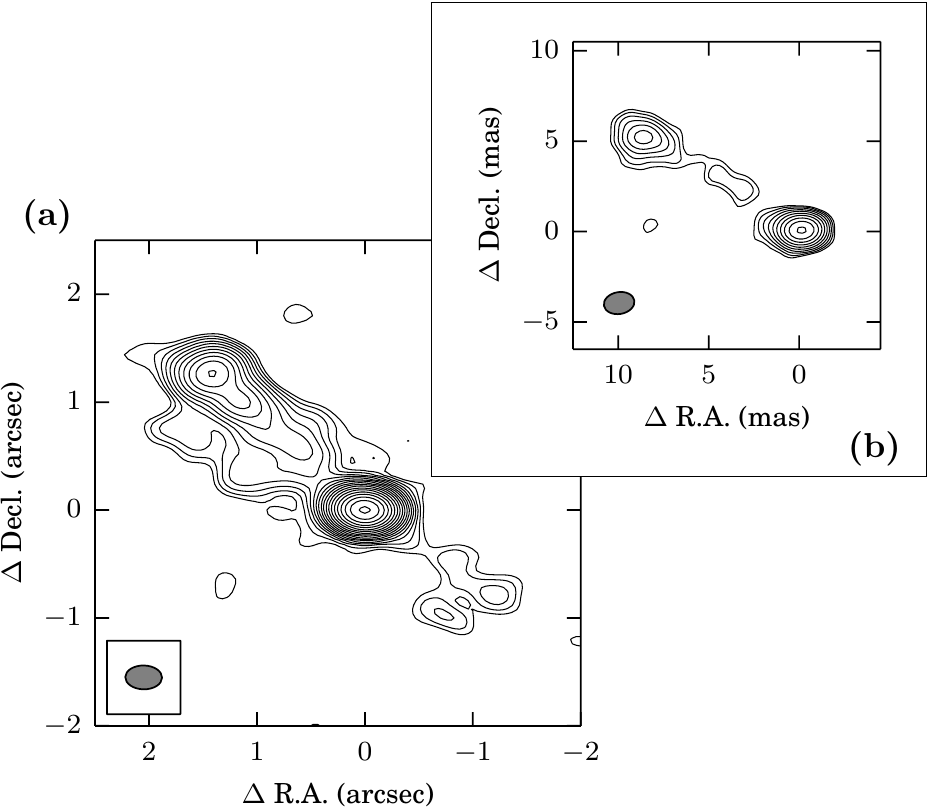}
  \caption{Naturally weighted maps of
    \sourcename{J1722+5654}. Contours increase by factors
    of$\sqrt{2}$. At this redshift,
    $1\arcsec{}=5.58~\mathrm{kpc}$. (a) VLA map (9~GHz). The first
    contour is at $21~\mathrm{\muup Jy\,beam^{-1}}$, about three times
    the map rms noise of $7~\mathrm{\muup Jy\,beam^{-1}}$, and the map
    peaks at $8.14~\mathrm{mJy\,beam^{-1}}$. The elliptical Gaussian
    restoring beam is shown (FWHM $0.341\arcsec\times0.221\arcsec$).
    (b) VLBA map (7.6~GHz) of the central bright feature. The first
    contour is at $0.2~\mathrm{mJy\,beam^{-1}}$, about three times the
    map rms noise of $0.07~\mathrm{mJy\,beam^{-1}}$, and the map peaks
    at $4.8~\mathrm{mJy\,beam^{-1}}$. The elliptical Gaussian
    restoring beam is shown (FWHM
    $1.71~\mathrm{mas}\times1.21~\mathrm{mas}$).}
  \label{fig:j1722+5654}
\end{figure}

On kiloparsec scales, \sourcename{J1722+5654}
(Figure~\ref{fig:j1722+5654}) contains a bright central core straddled
by a linear structure along PA 47\degr{}. The emission spans
4\farcs{}08 which corresponds to a 22.8~kpc projected length. The
parsec-scale structure is also linear, with three apparent emission
regions along PA 60\degr{}. The brightest parsec-scale emission is
from the SW component; if this is identified as the core, then the
parsec-scale morphology is a one-sided jet directed toward the
brighter kiloparsec-scale lobe. This component is 2.3 times brighter
than the NE component. We note that the structure could also be
interpreted as a two-sided jet with a faint core at the central
emission region, in which case the brighter ejected component is
directed toward the fainter kiloparsec-scale lobe. Misalignments as
large as 180\degr{} between the parsec- and kiloparsec-scale jets have
been seen~\citep{kharb_extended_2010}. However, the brightness
temperature of the SW component is more than 20~times higher than
either other component, which favors its identification as the
core~(see Section~\ref{sec:jet_speeds}).

\begin{deluxetable*}{c c c c c c c c c c c c c c c}
  \tablecaption{Measured properties.\label{table:computed}}
  \tablehead{
    \colhead{Source} &
    \colhead{$\ell_{\rm proj}$} &
    \colhead{$S_{\rm core,9}^{\rm kpc}$} &
    \colhead{$S_{\rm tot,9}^{\rm kpc}$} &
    \colhead{$\mathcal{I}_{9}^{\rm hs,j}$} &
    \colhead{$\alpha$} &
    \colhead{$L_{1.4}^{\rm ext}$} &
    \colhead{$R_{\rm hs}$} &
    \colhead{$\log R_{\rm v}$} &
    \colhead{$(\beta\cos\theta)_{\rm hs}$} &
    \colhead{$(\beta\cos\theta)_{x}$} &
    \colhead{Age} &
    \colhead{$S_{\rm core,7.6}^{\rm pc}$} &
    \colhead{$S_{\rm tot,7.6}^{\rm pc}$} &
    \colhead{$T_{\rm B}$} \\
    %
    \colhead{} &
    \colhead{(kpc)} &
    \colhead{(mJy)} &
    \colhead{(mJy)} &
    \colhead{(${\rm \muup{}Jy\, beam^{-1}}$)} &
    \colhead{} &
    \colhead{(W~Hz$^{-1}$)} &
    \colhead{} &
    \colhead{} &
    \colhead{} &
    \colhead{} &
    \colhead{($10^6$~y)} &
    \colhead{(mJy)} &
    \colhead{(mJy)} &
    \colhead{($10^{8}$~K)}\\
    \colhead{(1)} &
    \colhead{(2)} &
    \colhead{(3)} &
    \colhead{(4)} &
    \colhead{(5)} &
    \colhead{(6)} &
    \colhead{(7)} &
    \colhead{(8)} &
    \colhead{(9)} &
    \colhead{(10)} &
    \colhead{(11)} &
    \colhead{(12)} &
    \colhead{(13)} &
    \colhead{(14)} &
    \colhead{(15)}
    }
  \startdata
  J0953+2836 & 
  70.2 & 
  12.0 & 
  18.8 & 
  158 & 
  $-0.50$ & 
  25.9 & 
  1.9 & 
  1.94 & 
  0.082 & 
  0.028 & 
  14--22 & 
  4.35 &
  10.0 & 
  1.2  
  \\
  %
  J1435+3131 & 
  30.6 & 
  7.8 & 
  11.7 & 
  324 & 
  $-0.72$ & 
  25.4 & 
  2.2 & 
  1.90 & 
  0.098 & 
  0.15 & 
  1.3--2.0 & 
   5.1 &
  5.5 & 
   4.8  
  \\
  %
  J1722+5654 & 
  22.8 & 
  7.9 & 
  12.2 & 
  688 & 
  $-0.64$ & 
  25.3 & 
  13 & 
  1.52 & 
  0.31 & 
  0.20 & 
  0.6--1.0 & 
  5.4 &
  10.2 & 
  9.8  
  \enddata
  %
  \tablecomments{Columns are as follows: (1)~IAU name (J2000);
    (2)~projected full span of extended radio emission; (3)-(4)~flux
    density of the core component and the total integrated flux
    density from the 9~GHz VLA data; (5)~peak intensity of the 9~GHz
    emission in the approaching lobe; (6)~non-simultaneous 1.4--9~GHz
    spectral index; (7)~extended isotropic luminosity, extrapolated
    from 9~GHz to 1.4~GHz and $k$-corrected assuming $\alpha_{\rm
      ext}=-1$; (8)~ratio of peak intensity of approaching and
    receding lobes; (9)~ratio of optical to radio core luminosity,
    $k$-corrected to the emission frame assuming $\alpha_{\rm
      core}=0$, $\alpha_{\rm ext}=-1$, and $\alpha_{\rm
      optical}=-0.5$; (10)~projected jet speed computed from $R_{\rm
      hs}$; (11)~projected jet speed computed from jet/counter-jet
    length ratio; (12)~kinematic age of source implied by
    $(\beta\cos\theta)_x$ for $\theta=10$--$15\degr$; (13)~7.6~GHz
    VLBA core flux density; (14)~total 7.6~GHz VLBA flux density;
    (15)~emission frame VLBA core brightness temperature.}
\end{deluxetable*}

\section{Discussion}

\subsection{Morphologies}
The parsec- and kiloparsec-scale morphologies of these three sources
are similar to those found in other jetted AGN. Their kiloparsec scale
morphologies resemble those found in blazars, radio galaxies, and
other radio-loud NLS1s~\citep{cooper_mojave_2007, kharb_extended_2010,
  doi_kiloparsec-scale_2012}. Extended two-sided emission is present
in all three cases, with one lobe substantially brighter than the
other, presumably because it is approaching us. These resemble FR~II
radio galaxies, with brightened hotspots near the outer edges of the
kiloparsec-scale lobes evident at least on the approaching side. All
three sources are moderately core dominated at these scales, with
about 2/3 of the total emission in the kiloparsec core. The extended
isotropic luminosities $L_{1.4}^{\rm ext}$, computed by subtracting
the fitted core flux density from the total flux density and scaling
to 1.4~GHz assuming a spectral index $\alpha_{\rm ext}=-1$, lie in the
transition range between FR~I and FR~II galaxies, and fall along the
$L-z$ trend reported in \citet{kharb_extended_2010}.

On parsec scales, the sources are compatible with one-sided jet
morphologies. The parsec-scale core region in \sourcename{J1435+3131}
is especially blazar-like, consisting only of a compact core.  The
other two sources are compatible with an alternative interpretation as
a faint or undetected core with two-sided emission.

\subsection{Viewing Angles}
Two lines of reasoning suggest that the jet axes in these sources are
at relatively small angles to the line of sight.  First, the
  extended parsec-scale jets, where detected, are aligned with the
  kiloparsec-scale jets (this would not be true for
  \sourcename{J1722+5654} if it is misaligned by about 180\degr{}).
Misalignments due to projection effects are expected to be more common
for small viewing angles, so the alignment we find suggests the
viewing angles are not especially small, although our statistics are
limited~\citep{pearson_milliarcsecond_1988}.

Second, we reach a similar conclusion from $R_{\rm V}$, the ratio of
the observed radio core luminosity to the optical luminosity. The
latter acts as a proxy for the unbeamed, intrinsic emission. Motivated
by \citet{wills_improved_1995}, \citet{kharb_extended_2010} found
$R_{\rm V}$ to be a good indicator of viewing angle, where $\log
R_{\rm V}=\log L_{\rm core} - \log M_{\rm abs} - 13.69$ with $L_{\rm
  core}$ the 1.4~GHz core luminosity and $M_{\rm abs}$ the
$k$-corrected optical V-band magnitude. Assuming a flat core spectral
index, we find that the $\log R_{\rm V}$ values for these NLS1s
(1.5--1.9) are at the low end of the distribution for jet-selected
radio-loud AGN~\citep[e.g., the MOJAVE sample;][]{lister_mojave_2013,
  kharb_extended_2010}. This suggests the viewing angles of these
NLS1s are similar to the highest angles in the MOJAVE sources, roughly
10--15\degr{}~\citep{savolainen_relativistic_2010}.

\subsection{Jet Speeds}
\label{sec:jet_speeds}
Assuming the structure and evolution of these sources is intrinsically
symmetric, we can estimate the projected jet speed in two ways.  This
common assumption is supported in these particular cases by the
alignment of the approaching parsec-scale jet and the longer
kiloparsec-scale lobe~\citep{arshakian_asymmetric_2000}.  Both methods
have a degeneracy between the speed and the viewing angle $\theta$,
but assuming $\theta\gtrsim10\degr$, as seems to be the case,
$\beta\cos\theta\sim\beta$ to within the likely accuracy of our
measurements. First, we consider the flux ratio between the hotspots
in the approaching and receding jets. If absorption is negligible
along the additional optical path to the counter-jet, the expected
ratio is
\begin{equation}
  \label{eq:jet-counterjet-ratio}
  R_{\rm hs}=\left(\frac{1+(\beta\cos\theta)_{\rm hs}}{1-(\beta\cos\theta)_{\rm hs}}\right)^{3-\alpha},
\end{equation}
where $(\beta\cos\theta)_{\rm hs}$ is the projected speed at which the
lobes advance at angle $\theta$ to the line of sight.  For this
calculation, we use the ratio of the peak intensity (in $\rm\muup Jy\
beam^{-1}$) between the jet and counter-jet lobes, ($R_{\rm
  hs}=\mathcal{I}_{9}^{\rm hs,j}/\mathcal{I}_{9}^{\rm hs,cj}$). We
obtain similar results if we instead use the ratio between the flux
densities obtained by fitting a Gaussian component model. For
\sourcename{J0953+2836} and \sourcename{J1722+5654}, if we interpret
the parsec-scale jets as two-sided, the flux density ratios of 1.2 and
2.3 imply jet speeds $\beta\cos\theta$=0.04 and 0.18, comparable to
those found at kiloparsec scales.

We can also examine the asymmetry in length between the jet and
counter-jet. If the structure is intrinsically symmetric, differential
light travel time will cause the approaching jet to appear longer than
the receding jet~\citep{banhatti_expansion_1980}. In this case, the ratio $x\equiv(r_{\rm
  j}-r_{\rm cj})/(r_{\rm j}+r_{\rm cj})=(\beta\cos\theta)_{x}$ where
$r_{\rm j}$ and $r_{\rm cj}$ are the lengths of the jet and
counter-jet, and $(\beta\cos\theta)_{x}$ is the jet speed estimate.
This method gives comparable results to the beaming ratio method, so
overall we find $\beta\cos\theta\sim0.03$--0.3. These correspond to
Doppler factors in the range $\delta=1$--1.5, where
$\delta=(1-\beta^2)^{1/2}/(1-\beta\cos\theta)$.

Doppler beaming will increase the apparent brightness temperature of
an emitting region relative to its intrinsic brightness temperature,
$T_{\rm B,int}$, with the observed $T_{\rm B}=\delta T_{\rm
  B,int}$. Using our VLBA data, we computed $T_{\rm B}$ in the
emission frame from an elliptical Gaussian model as $T_{\rm
  B}=(1.22\times10^{12}~{\rm K})\times S(1+z)/(a_{\rm maj}a_{\rm
  min}\nu^{2})$ where $\nu$ is the observing frequency in GHz and $S$,
$a_{\rm maj}$, and $a_{\rm min}$ are the flux density in janskys and
the major and minor axis FWHMs (in mas) of the core component. The
brightness temperatures we find are all substantially below the
equipartition limit of about $10^{11}$~K, compatible with a low
Doppler factor~\citep{readhead_equipartition_1994}. In each source,
the presumed core has the highest brightness temperature of all fitted
components by at least a factor of eight.

The kinematic ages of these sources, computed assuming the 
  approaching hotspots advanced outward steadily at
  their current speeds ($\beta\sim0.03-0.3$) at a viewing angle
  $\theta\sim10$--15\degr{}, are found to be a few times
    $10^{5}$--$10^{7}$~y. These are in the range found for other
  extended radio-loud NLS1s, consistent with NLS1s being much younger
  than blazars or radio galaxies~\citep{doi_kiloparsec-scale_2012}.

\section{Conclusions}
Extended emission on kiloparsec scales was detected in all three of
our targets. This is perhaps surprising in light of the small number
of spatially extended NLS1s previously known. However, few have been
observed with the requisite angular resolution and sensitivity to
detect the extensions we found here. Because our targets were
arbitrarily selected from a parent sample of bright, radio-loud NLS1s,
there was no reason all three were especially likely to show extended
emission. Thus, our findings suggest that other NLS1s in the parent
sample are also likely to be extended.  Observations of a larger
sample are needed to verify this and, if true, to investigate whether
this is generally true of all radio-loud NLS1s or is a peculiar
property of the parent sample.  The morphologies of these sources on
both kiloparsec and parsec scales are consistent with the presence of
a mildly relativistic jet at a moderately small (10--15\degr{}) angle
to the line of sight. The three sources we observed and most
previously known extended NLS1s where resolution permitted
classification show FR~II-like edge-brightened lobes. In contrast,
\citet{doi_fanaroff-riley_2014} recently reported edge-darkened
FR~I-like extended radio emission in the NLS1 \sourcename{Mrk
  1239}. The latter is radio quiet, while the others are radio loud,
suggesting that radio loudness may be connected to intrinsic source
properties, such as jet power, rather than orientation effects.

\acknowledgements The authors thank T.~G.~Arshakian for helpful
suggestions and E.~Angelakis and the F-GAMMA program for sharing
the preliminary data we used to check our VLBA calibration. This work
was supported by the National Aeronautics and Space Administration
(NASA) through \emph{Fermi} Guest Investigator grant NNX13AO79G. The
National Radio Astronomy Observatory is a facility of the National
Science Foundation operated under cooperative agreement by Associated
Universities, Inc. This work made use of the Swinburne University of
Technology software correlator, developed as part of the Australian
Major National Research Facilities Programme and operated under
licence. This research has made use of the NASA/IPAC Extragalactic
Database (NED) which is operated by the Jet Propulsion Laboratory,
California Institute of Technology, under contract with the National
Aeronautics and Space Administration. This work made use of Ned
Wright's Javascript Cosmology
Calculator~\citep{wright_cosmology_2006}.

\end{document}